\documentclass[prb,preprintnumbers,amsmath,amssymb,floatfix]{revtex4}

\usepackage{graphicx}
\usepackage{subfigure}
\usepackage{dcolumn}

\begin{document}

\title{Quantum theory of light double-slit diffraction}
\author{Xiang-Yao Wu$^{a}$ \footnote{E-mail: wuxy2066@163.com},
 Hong Li$^{a}$, Bo-Jun Zhang$^{a}$, Ji Ma$^{a}$, Xiao-Jing Liu$^{a}$
Nuo Ba$^{a}$\\ He Dong$^{a}$, Si-Qi Zhang$^{a}$, Jing Wang$^{a}$,
Yi-Heng Wu$^{b}$ and Xin-Guo Yin$^{c}$ \footnote{E-mail:
msyyxg3081@126.com} }
\affiliation{a.Institute of Physics, Jilin Normal University, Siping 136000 \\
b. Institute of Physics, Jilin University, Changchun 130012
China\\
c. Institute of Physics, Huaibei Normal University, Huaibei 235000
 }

\begin{abstract}

In this paper, we study the light double-slit diffraction
experiment with quantum theory approach. Firstly, we calculate the
light wave function in slits by quantum theory of photon.
Secondly, we calculate the diffraction wave function with
Kirchhoff's law. Thirdly, we give the diffraction intensity of
light double-slit diffraction, which is proportional to the square
of diffraction wave function. Finally, we compare calculation
result of quantum theory and classical electromagnetic theory with
the experimental data. We find the quantum calculate result is
accordance with the experiment data, and the classical calculation
result with certain deviation. So, the quantum theory is more
accurately approach for studying light diffraction.
 \\
\vskip 5pt


\end{abstract}
\maketitle

\maketitle {\bf 1. Introduction} \vskip 8pt

In recent years, quantum information science has advanced rapidly,
both at the level of fundamental research and technological
development. For instance, quantum cryptography systems have
become commercially available [1]. Classical optical lithography
technology is facing its limit due to the diffraction effect of
light. It is known that the nonclassical phenomena of two photon
interference and two- photon ghost diffraction and imaging, have
classical counterparts [2-3]. Two photon interference of classical
light has been first discovered in the pioneering experiments by
Hanbury Brown and Twiss and since then was observed with various
sources, including true thermal ones, and coherent ones [4-7].
Somewhat later, ghost imaging with classical light has been
demonstrated, both in the near-field and far-field domains [8-10].
The present optical imaging technologies, such as optical
lithography, have reached a spatial resolution in the
sub-micrometer range, which comes up against the diffraction limit
due to the wavelength of light. However, the guiding principle of
such technology is still based on the classical diffraction theory
established by Fresnel, Kirchhoff and others more than a hundred
years ago. Recently, the use of quantum- correlated photon pairs
to overcome the classical diffraction limit was proposed and
attracted much attention. Obviously, quantum theory approaches are
necessary to explain the diffraction-interference of the
quantum-correlated multi photon state. As is well known, the
classical optics with its standard wave- theoretical methods and
approximations, such as Huygens' and Kirchhoff's theory, has been
successfully applied to classical optics, and has yielded good
agreement with many experiments. However, light interference and
diffraction are quantum phenomena, and its full description needs
quantum theory approach. In 1924, Epstein and Ehrenfest had
firstly studied light diffraction with the old quantum theory,
i.e., the quantum mechanics of correspondence principle, and
obtained a identical result with the classical optics [11-17]. In
this paper, we study the double-slit diffraction of light with the
approach of relativistic quantum theory of photon. In view of
quantum theory, the light has the nature of wave, and the wave is
described by wave function. We calculate the light wave function
in slits by quantum theory of photon, where the diffraction wave
function can be calculated by the Kirchhoff's law. The diffraction
intensity is proportional to the square of diffraction wave
function. We can obtain the diffraction intensity by calculating
the light wave function distributing on display screen. We compare
calculation results of quantum theory and classical
electromagnetic theory with the experimental data. When the
decoherence effects are considered, we find the quantum calculate
result is in accordance with the experiment data, but the
classical calculation result with certain deviation. In order to
study the light double-slit diffraction more accurately, it should
be applied the new approach of quantum theory.

 \vskip 5pt
 \setlength{\unitlength}{0.1in}
\begin{center}
\begin{figure}[tbp]
\includegraphics[width=8.5 cm]{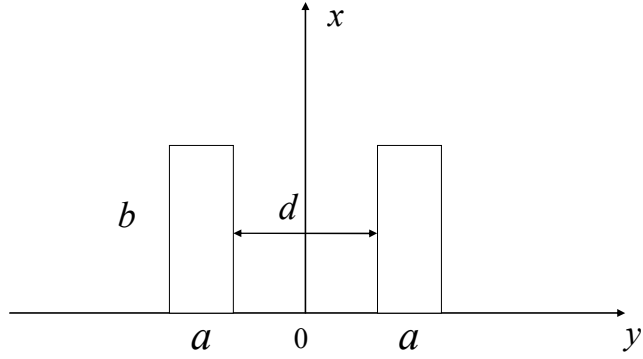}
\caption{Light double-slit diffraction}
\end{figure}
\end{center}
{\bf 2. Quantum approach of light diffraction}

\vskip 8pt In an infinite plane, we consider a double-slit, its
width $a$, length $b$ and the slit-to-slit distance $d$ are shown
in Fig.\,1. The $x$ axis is along the slit length $b$ and the $y$
axis is along the slit width $a$, We calculate the light wave
function in the left slit with the light of the relativistic wave
equation. At time $t$, we suppose that the incoming plane wave
travels along the $z$ axis. It is
\begin{eqnarray}
\vec{\psi}_{0}(z,
t)&=&\vec{A}e^{\frac{i}{\hbar}(pz-Et)}\nonumber\\
&=&\sum_{j}A_{j}\cdot e^{\frac{i}{\hbar}(pz-Et)}\vec{e}_{j}\nonumber\\
&=&\sum_{j}\psi_{0j}\cdot e^{-\frac{i}{\hbar}Et}\vec{e}_{j},
\end{eqnarray}
where $\psi_{0j}=A_{j}\cdot e^{\frac{i}{\hbar}pz}$,  $j= x, y, z$
and $\vec{A}$ is a constant vector. The time-dependent
relativistic wave equation of light is [12]
\begin{equation}
i\hbar\frac{\partial}{\partial
t}\vec{\psi}(\vec{r},t)=c\hbar\nabla\times\vec{\psi}(\vec{r},t)+V\vec{\psi}(\vec{r},t),
\end{equation}
where $c$ is light velocity. From Eq. (2), we can find the light
wave function $\vec{\psi}(\vec{r},t)\rightarrow 0$  when
$V(\vec{r})\rightarrow\infty$. The potential energy of light in
the left slit is
\begin{eqnarray}
V(x,y,z)= \left \{ \begin{array}{ll}
   0  \hspace{0.3in} 0\leq x\leq b, -\frac{d}{2}-a\leq y\leq -\frac{d}{2}, 0\leq z\leq c', \\
   \infty  \hspace{0.3in}  otherwise,
   \end{array}
   \right.
\end{eqnarray}
where $c'$ is the slit thickness. We can get the time-dependent
relativistic wave equation in the slit ($V(x,y,z)=0$), it is
\begin{equation}
i\hbar\frac{\partial}{\partial
t}\vec{\psi_1}(\vec{r},t)=c\hbar\nabla\times\vec{\psi_1}(\vec{r},t),
\end{equation}
by derivation on  Eq. (4) about the time t and multiplying
$i\hbar$ both sides, we have
\begin{equation}
(i\hbar)^2\frac{\partial^2}{\partial
t^2}\vec{\psi_1}(\vec{r},t)=c\hbar\nabla\times
i\hbar\frac{\partial}{\partial t}\vec{\psi_1}(\vec{r},t),
\end{equation}
substituting Eq. (4) into (5), we have
\begin{eqnarray}
\frac{\partial^2}{\partial t^2}\vec{\psi_1}(\vec{r},t)&
=&-c^2[\nabla(\nabla\cdot\vec{\psi_1}(\vec{r},t))-\nabla^2\vec{\psi_1}(\vec{r},t)],
\end{eqnarray}
where the formula $\nabla\times\nabla\times
\vec{B}=\nabla(\nabla\cdot\vec{B})-\nabla^2\vec{B}$. From Ref.
[11], the photon wave function is
$\vec{\psi_1}(\vec{r},t)=\sqrt{\frac{\varepsilon_{0}}{2}}(\vec{E}(\vec{r},t)+i\sigma
c\vec{B}(\vec{r},t))$, we have
\begin{equation}
\nabla\cdot\vec{\psi_1}(\vec{r},t)=0,
\end{equation}
from Eq. (6) and (7), we have
\begin{equation}
(\frac{\partial^2}{\partial
t^2}-c^2\nabla^2)\vec{\psi_1}(\vec{r},t)=0.
\end{equation}
The Eq. (8) is the same as the classical wave equation of light.
Here, it is a quantum wave equation of light, since it is obtained
from the relativistic wave equation (2), and it satisfied the new
quantum boundary condition: when
$\vec{\psi_1}(\vec{r},t)\rightarrow 0$,
$V(\vec{r})\rightarrow\infty$. It is different from the classic
boundary condition. When the photon wave function
$\vec{\psi_1}(\vec{r},t)$ change with determinate frequency
$\omega$, the wave function of photon can be written as
\begin{equation}
\vec{\psi_1}(\vec{r},t)=\vec{\psi_1}(\vec{r})e^{-i\omega t},
\end{equation}
substituting Eq. (9) into (8), we can get
\begin{equation}
\frac{\partial^{2}\vec{\psi_1}(\vec{r})}{\partial
x^{2}}+\frac{\partial^{2}\vec{\psi_1}(\vec{r})}{\partial
y^{2}}+\frac{\partial^{2}\vec{\psi_1}(\vec{r})}{\partial
z^{2}}+\frac{4\pi^{2}}{\\\lambda^{2}}\vec{\psi_1}(\vec{r})=0,
\end{equation}
and the wave function satisfies boundary conditions
\begin{equation}
\psi_1(0,y,z)=\psi_1(b,y,z)=0,
\end{equation}
\begin{equation}
\psi_1(x,-\frac{d}{2}-a,z)=\psi_1(x,-\frac{d}{2},z)=0,
\end{equation}
the photon wave function $\vec{\psi}(\vec{r})$ can be wrote
\begin{eqnarray}
\vec{\psi_1}(\vec{r})&=&\psi_{1x}(\vec{r})\vec{e}_{x}+\psi_{1y}(\vec{r})\vec{e}_{y}+\psi_{1z}(\vec{r})\vec{e}_{z}\nonumber\\
&=&\sum_{j=x,y,z}\psi_{1j}(\vec{r})\vec{e}_{j},
\end{eqnarray}
where $j$ is $x$, $y$ or $z$. Substituting Eq. (13) into (10),
(11) and (12), we have the component equation
\begin{equation}
\frac{\partial^{2}\psi_{1j}(\vec{r})}{\partial
x^{2}}+\frac{\partial^{2}\psi_{1j}(\vec{r})}{\partial
y^{2}}+\frac{\partial^{2}\psi_{1j}(\vec{r})}{\partial
z^{2}}+\frac{4\pi^{2}}{\\\lambda^{2}}\psi_{1j}(\vec{r})=0,
\end{equation}
\begin{equation}
\psi_{1j}(0,y,z)=\psi_{1j}(b,y,z)=0,
\end{equation}
\begin{equation}
\psi_{1j}(x,-\frac{d}{2}-a,z)=\psi_{1j}(x,-\frac{d}{2},z)=0,
\end{equation}
the partial differential equation (14) can be solved by the method
of separation of variable. By writing
\begin{equation}
\psi_{1j}(x,y,z)=X_{1j}(x)Y_{1j}(y)Z_{1j}(z).
\end{equation}
From Eqs. (14-17), we can get the general solution of Eq. (14)
\begin{equation}
\psi_{1j}(x,y,z)=\sum_{mn}\sin\frac{n\pi}{b}x\cdot(D_{mnj}\cos\frac{m\pi}{a}y+D'_{mnj}\sin\frac{m\pi}{a}y)\cdot\exp[i\sqrt{\frac{4\pi^{2}}{\lambda^{2}}-(\frac{m\pi}{a})^{2}-(\frac{n\pi}{b})^{2}}\cdot
z],
\end{equation}
since the wave functions are continuous at $z=0$, we have
\begin{equation}
\vec{\psi}_{0}(x,y,z;t)\mid_{z=0}=\vec{\psi_{1}}(x,y,z;t)\mid_{z=0},
\end{equation}
or, equivalently,
\begin{eqnarray}
\psi_{0j}(x,y,z)\mid_{z=0}&=&\psi_{1j}(x,y,z)\mid_{z=0}.\hspace{0.3in}(j=x,y,z)
\end{eqnarray}
From Eq. (1), (18) and (20), we obtain the coefficient $D_{mnj}$
by fourier transform
\begin{eqnarray}
D_{mnj}&=&\frac{4}{a\cdot
b}\int^{b}_{0}\int^{-\frac{d}{2}}_{-\frac{d}{2}-a}A_{1j}\cdot\sin\frac{n\pi}{b}x\cdot\cos\frac{m\pi}{a}yd_{x}d_{y}\nonumber\\
&=&\frac{-16A_{1j}}{(2m+1)\cdot(2n+1)\cdot\pi^{2}}\sin\frac{(2m+1)\cdot\pi}{2a}\cdot
d,
\end{eqnarray}
\begin{eqnarray}
D'_{mnj}&=&\frac{4}{a\cdot
b}\int^{b}_{0}\int^{-\frac{d}{2}}_{-\frac{d}{2}-a}A_{1j}\cdot\sin\frac{n\pi}{b}x\cdot\sin\frac{m\pi}{a}yd_{x}d_{y}\nonumber\\
&=&\frac{16A_{1j}}{(2m+1)\cdot(2n+1)\cdot\pi^{2}}\cos\frac{(2m+1)\cdot\pi}{2a}\cdot
d,
\end{eqnarray}
substituting Eq. (21) and (22) into (18), we have
\begin{eqnarray}
\psi_{1j}(x,y,z)&=&\sum_{j=x,y,z}\sum^{\infty}_{m,n=0}\frac{-16A_{1j}}{(2m+1)\cdot(2n+1)\cdot\pi^{2}}\cdot\sin\frac{(2n+1)\pi}{b}x\nonumber\\&&
\cdot [\sin\frac{(2m+1)\cdot\pi
d}{2a}\cdot\cos\frac{(2m+1)\cdot\pi}{a}y+\cos\frac{(2m+1)\cdot\pi
d}{2a}\cdot\sin\frac{(2m+1)\cdot\pi}{a}y]\nonumber\\&&\exp[i\sqrt{\frac{4\pi^{2}}{\lambda^{2}}-(\frac{(2m+1)\pi}{a})^{2}-(\frac{(2n+1)\pi}{b})^{2}}\cdot
z],
\end{eqnarray}
substituting Eq. (23) into (9) and (13), we can obtain the photon
wave function $\vec{\psi}_{1}(x,y,z,t)$ in slit
\begin{eqnarray}
\vec{\psi}_{1}(x,y,z,t)&=&\sum_{j=x,y,z}\psi_{1j}(x,y,z,t)\vec{e}_{j}\nonumber\\
&=&\sum_{j=x,y,z}\sum^{\infty}_{m,n=0}\frac{-16A_{1j}}{(2m+1)\cdot(2n+1)\cdot\pi^{2}}
\sin\frac{(2n+1)\pi}{b}x\nonumber\\&&\cdot
[\sin\frac{(2m+1)\cdot\pi
d}{2a}\cdot\cos\frac{(2m+1)\cdot\pi}{a}y+\cos\frac{(2m+1)\cdot\pi
d}{2a}\cdot\sin\frac{(2m+1)\cdot\pi}{a}y]\cdot\nonumber\\&&
\exp[i\sqrt{\frac{4\pi^{2}}{\lambda^{2}}-(\frac{(2m+1)\pi}{a})^{2}-(\frac{(2n+1)\pi}{b})^{2}}\cdot
z]\cdot\exp[-i\omega t]\vec{e_{j}}.
\end{eqnarray}
The potential energy of light in the right slit is
\begin{eqnarray}
V(x,y,z)= \left \{ \begin{array}{ll}
   0  \hspace{0.3in} 0\leq x\leq b,\frac{d}{2}\leq y\leq \frac{d}{2}+a, 0\leq z\leq c', \\
   \infty  \hspace{0.3in}  otherwise,
   \end{array}
   \right.
\end{eqnarray}
and the wave function satisfies boundary conditions
\begin{equation}
\psi_2(0,y,z)=\psi_2(b,y,z)=0,
\end{equation}
\begin{equation}
\psi_2(x,\frac{d}{2},z)=\psi_2(x,\frac{d}{2}+a,z)=0,
\end{equation}
similarly, we can obtain the light wave function
$\vec{\psi}_{2}(x,y,z,t)$ in the right slit
\begin{eqnarray}
\vec{\psi}_{2}(x,y,z,t)&=&\sum_{j=x,y,z}\sum^{\infty}_{m,n=0}\frac{-16A_{2j}}{(2m+1)\cdot(2n+1)\cdot\pi^{2}}
\sin\frac{(2n+1)\pi}{b}x\nonumber\\&&\cdot
[\sin\frac{(2m+1)\cdot\pi
d}{2a}\cdot\cos\frac{(2m+1)\cdot\pi}{a}y-\cos\frac{(2m+1)\cdot\pi
d}{2a}\cdot\sin\frac{(2m+1)\cdot\pi}{a}y]\cdot\nonumber\\&&
\exp[i\sqrt{\frac{4\pi^{2}}{\lambda^{2}}-(\frac{(2m+1)\pi}{a})^{2}-(\frac{(2n+1)\pi}{b})^{2}}\cdot
z]\cdot\exp[-i\omega t]\vec{e_{j}}.
\end{eqnarray}
{\bf 3. The wave function of light diffraction} \vskip 8pt In the
section 2, we have calculated the photon wave function in slit. In
the following, we will calculate diffraction wave function. we can
calculate the wave function in the diffraction area. From the slit
wave function component $\psi_{j}(\vec{r},t)$, we can calculate
its diffraction wave function component $\Phi_{j}(\vec{r},t)$ by
Kirchhoff's law. It can be calculated by the formula[13]
\begin{equation}
\Phi_{j}(\vec{r},t)=-\frac{1}{4\pi}\int_{s_{0}}\frac{e^{ikr}}{r}\vec{n}\cdot[\bigtriangledown^{'}\psi_{j}
+(ik-\frac{1}{r})\frac{\vec{r}}{r}\psi_{j}]ds,
\end{equation}
the total diffraction wave function is
\begin{eqnarray}
\vec{\Phi}(\vec{r},t)&=&\sum_{j=x,y,z}\Phi_{j}(\vec{r},t)\vec{e}_{j},
\end{eqnarray}
in the following, we firstly calculate the diffraction wave
function of the top slit, it is
\begin{equation}
\Phi_{1j}(\vec{r}_1,t)=-\frac{1}{4\pi}\int_{s_{1}}\frac{e^{ikr_1}}{r_1}
{ \vec{n}}\cdot[\nabla'\psi_{1j} +(ik-\frac{1}{r_1})\frac{
\vec{r}_1}{r_1}\psi_{1j}]ds.
\end{equation}
The diffraction area is shown in Fig. 2, where
$k=\frac{2\pi}{\lambda}$, $s_{1}$ is the area of the top slit,
${\vec{r'_1}}$ is the position of a point on the surface (z=c),
$P$ is an arbitrary point in the diffraction area, and ${\vec n}$
is a unit vector, which is normal to the surface of the slit.
 \vskip 5pt
 \setlength{\unitlength}{0.1in}
\begin{center}
\begin{figure}[tbp]
\includegraphics[width=6.5 cm]{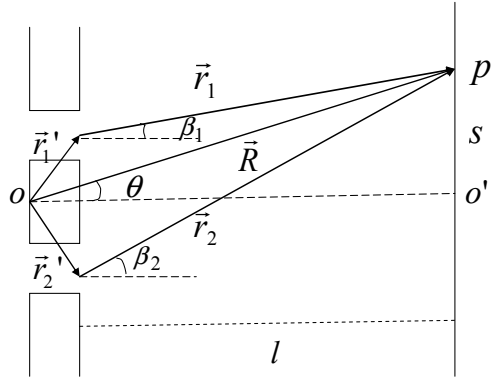}
\caption{Di®raction area of the double slits}
\end{figure}
\end{center}
In Fig.\,2, we firstly consider the up slit, there are
\begin{eqnarray}
r_1&=&R-\frac{\vec{R}}{R}\cdot{\vec{r'}_1}
\approx R-\frac{{\vec{r_1}}}{r_1}\cdot{\vec{r'_1}}\nonumber\\
 &=&R-\frac{\vec{k_1}}{k}\cdot{\vec{r'_1}},
\end{eqnarray}
and then,
\begin{eqnarray}
\frac{e^{ikr_1}}{r_1}&=&\frac{e^{ik(R-\frac{{\vec
r_1}}{r}\cdot{\vec r'_1})}} {R-\frac{{\vec r_1}}{r_1}\cdot{\vec
r'_1}} =\frac{e^{ikR}e^{-i{\vec k_{1}}\cdot{\vec r'_1}}}
{R-\frac{{\vec r_1}}{r}\cdot{\vec r'_1}}\nonumber\\
&\approx&\frac{{e^{ikR}e^{-i{\vec k_{1}}\cdot{\vec r'_1}}}}{R}
\hspace{0.3in}(|{\vec r'_1}|\ll R),
\end{eqnarray}
with ${\vec{k}_{1}}=k\vec{r}_1/{r_1}$. Substituting Eq. (32) and
(33) into Eq. (31), we can obtain
\begin{eqnarray}
\Phi_{1j}(x,y,z;t)&&=-\frac{e^{ikR}}{4\pi R}e^{-i\omega
t}\sum_{m=0}^{\infty}\sum_{n=0}^{\infty}\frac{16A_{1j}}{(2m+1)(2n+1)\pi^{2}}
\exp[i\sqrt{\frac{4\pi^{2}}{\lambda^{2}}-(\frac{(2m+1)\pi}{a})^{2}-(\frac{(2n+1)\pi}{b})^{2}}\cdot
c'\nonumber\\&&[i\sqrt{\frac{4\pi^{2}}{\lambda^{2}}-(\frac{(2m+1)\pi}{a})^{2}-(\frac{(2n+1)\pi}{b})^{2}}+i\vec
n\cdot \vec k_{1}-\frac{\vec n\cdot \vec
R}{R^{2}}]\nonumber\\&&\int_{s_{1}}\exp[-i\vec k_{1}\cdot \vec
r']\cdot [\sin\frac{(2m+1)\pi d}{2a}\cos\frac{(2m+1)\pi
y}{a}+\cos\frac{2m+1)\pi d}{2a}\sin\frac{2m+1)\pi y}{a}]dx dy.
\end{eqnarray}
For the second diffraction slit, we assume the angle between
${\vec k_{1}}$ and $x$ axis ($y$ axis) is $\frac{\pi}{2}-\alpha$
($\frac{\pi}{2}-\beta_1$), and $\alpha (\beta_1)$ is the angle
between ${\vec k_{1}}$ and the surface of $yz$ ($xz$), then we
have
\begin{eqnarray}
k_{1x}=k\sin \alpha,\hspace{0.3in} k_{1y}=k\sin \beta_1,
\end{eqnarray}
\begin{eqnarray}
{\vec n}\cdot {\vec k_{1}}=k\cos \theta,
\end{eqnarray}
where $\theta$ is the angle between ${\vec k_{1}}$ and $z$ axis,
and the angles $\theta$, $\alpha$, $\beta_1$ satisfy the equation
\begin{equation}
\cos^{2}\theta+\cos^{2}(\frac{\pi}{2}-\alpha)+\cos^{2}(\frac{\pi}{2}-\beta_1)=1,
\end{equation}
with $R=\sqrt{l^{2}+s^{2}}$. Substituting Eqs. (35)-(37) into Eq.
(34) yields
\begin{eqnarray}
\Phi_{1j}(x,y,z;t)&=&-\frac{e^{ikR}}{4\pi R}e^{-i\omega
t}e^{-ik\cos\theta\cdot
c'}\sum_{m=0}^{\infty}\sum_{n=0}^{\infty}\frac{16A_{1j}}{(2m+1)(2n+1)\pi^2}
e^{i\sqrt{\frac{4\pi^{2}}{\lambda^{2}}-(\frac{(2n+1)\pi}{b})^{2}-(\frac{(2m+1)\pi}{a})^{2}}\cdot
c'}\nonumber\\&&
[i\sqrt{\frac{4\pi^{2}}{\lambda^{2}}-(\frac{(2n+1)\pi}{b})^{2}-(\frac{(2m+1)\pi}{a})^{2}}+(ik-\frac{1}{R})\cdot\sqrt{\cos^{2}\alpha-\sin^{2}\beta_{1}}]\nonumber\\&&
\int^{b}_{0}e^{-ik\sin\alpha\cdot
x}\sin\frac{(2n+1)\pi}{b}xdx\int^{-\frac{d}{2}}_{-\frac{d}{2}-a}e^{-ik\sin\beta_{1}\cdot
y} \sin \frac{(2m+1)\pi}{a}(\frac{d}{2}+y)dy.
\end{eqnarray}
Substituting Eq. (38) into (30), we can get the diffraction
function of the up slit
\begin{eqnarray}
\vec{\Phi_{1}}(x,y,z;t)&=&\sum_{j=x,y,z}\Phi_{1j}(x,y,z;t)\vec{e}_{j}\nonumber\\&=&-\frac{e^{ikR}}{4\pi
R}e^{-i\omega t}e^{-ik\cos\theta\cdot
c'}\sum_{j=x,y,z}\sum_{m=0}^{\infty}\sum_{n=0}^{\infty}\frac{16A_{1j}}{(2m+1)(2n+1)\pi^2}
e^{i\sqrt{\frac{4\pi^{2}}{\lambda^{2}}-(\frac{(2n+1)\pi}{b})^{2}-(\frac{(2m+1)\pi}{a})^{2}}\cdot
c'}\nonumber\\&&
[i\sqrt{\frac{4\pi^{2}}{\lambda^{2}}-(\frac{(2n+1)\pi}{b})^{2}-(\frac{(2m+1)\pi}{a})^{2}}+(ik-\frac{1}{R})\cdot\sqrt{\cos^{2}\alpha-\sin^{2}\beta_{1}}]\nonumber\\&&
\int^{b}_{0}e^{-ik\sin\alpha\cdot
x}\sin\frac{(2n+1)\pi}{b}xdx\int^{-\frac{d}{2}}_{-\frac{d}{2}-a}e^{-ik\sin\beta_{1}\cdot
y} \sin \frac{(2m+1)\pi}{a}(\frac{d}{2}+y)dy\vec{e}_{j}.
\end{eqnarray}
Similarly, the diffraction wave function of the down slit is
\begin{eqnarray}
\vec{\Phi_{2}}(x,y,z;t)&=&-\frac{e^{ikR}}{4\pi R}e^{-i\omega
t}e^{-ik\cos\theta\cdot
c'}\sum_{j=x,y,z}\sum_{m=0}^{\infty}\sum_{n=0}^{\infty}\frac{16A_{2j}}{(2m+1)(2n+1)\pi^2}
e^{i\sqrt{\frac{4\pi^{2}}{\lambda^{2}}-(\frac{(2n+1)\pi}{b})^{2}-(\frac{(2m+1)\pi}{a})^{2}}\cdot
c'}\nonumber\\&&
[i\sqrt{\frac{4\pi^{2}}{\lambda^{2}}-(\frac{(2n+1)\pi}{b})^{2}-(\frac{(2m+1)\pi}{a})^{2}}+(ik-\frac{1}{R})\cdot\sqrt{\cos^{2}\alpha-\sin^{2}\beta_{2}}]\nonumber\\&&
\int^{b}_{0}e^{-ik\sin\alpha\cdot
x}\sin\frac{(2n+1)\pi}{b}xdx\int^{\frac{d}{2}+a}_{\frac{d}{2}}e^{-ik\sin\beta_{2}\cdot
y} \sin \frac{(2m+1)\pi}{a}(\frac{d}{2}-y)dy\vec{e}_{j},
\end{eqnarray}
where $d$ is the two slit distance. The total diffraction wave
function for the double-slit is
\begin{eqnarray}
\vec{\Phi}(x,y,z;t)=c_{1}\vec{\Phi_{1}}(x,y,z;t)+c_{2}\vec{\Phi_{2}}(x,y,z;t).
\end{eqnarray}
where $c_{1}$ and $c_{2}$ are superposition coefficients, and
satisfy the equation
\begin{equation}
|c_{1}^{2}|+|c_{2}^{2}|=1.
\end{equation}
For the double-slit diffraction, we can obtain the relative
diffraction intensity $I$ on the display screen
\begin{equation}
I\propto|\vec{\Phi}(x,y,z;t)|^{2}.
\end{equation}
\vskip 8pt {\bf 4. The relative diffraction intensity $I$ on the
display screen} \vskip 8pt Decoherence is introduced here using a
simple phenomenological theoretical model that assumes an
exponential damping of the interferences [19], i.e., the
decoherence is the dynamic suppression of the interference terms
owing to the interaction between system and environment. Eq. (41)
describes the coherence state coherence superposition, without
considering the interaction of system with external environment.
When we consider the effect of external environment, the total
wave function of system and environment for the double-slit
factorizes as [19]
\begin{eqnarray}
\vec{\Phi}(x,y,z;t)=c_{1}\vec{\Phi_{1}}(x,y,z;t)\otimes|E_{1}>_{t}+c_{2}\vec{\Phi_{2}}(x,y,z;t)\otimes|E_{2}>_{t}.
\end{eqnarray}
where $\otimes|E_{1}>_{t}$ and $\otimes|E_{2}>_{t}$ describe the
state of the environment. Now, the diffraction intensity on the
screen is given by [19]
\begin{equation}
I=(1+|\alpha_{t}|^{2})[c_{1}^{2}|\vec{\Phi_{1}}|^{2}+c_{2}^{2}|\vec{\Phi_{2}}
|^{2}+2c_{1}c_{2}\Lambda_{t}R{e}(\vec{\Phi_{1}^{^{*}}}+
\vec{\Phi_{2}})],
\end{equation}
where $\alpha_{t}=_{t}<E_{2}|E_{1}>_{t}$, and
$\Lambda_{t}=\frac{2|\alpha_{t}|^{2}}{1+|\alpha_{t}|^{2}}$. Thus,
$\Lambda_{t}$ is defined as the quantum coherence degree. The
fringe visibility of n is defined as [19]
\begin{equation}
v=\frac{I_{max}-I_{min}}{I_{max}+I_{min}},
\end{equation}
where $I_{max}$ and $I_{min}$ are the intensities corresponding to
the central maximum and the first minimum next to it,
respectively. The value for the fringe visibility of $\nu=0.873$
is obtained in the experiment [18], and the quantum coherence
degree $\Lambda_{t}\simeq v$ [19]. Eq. (45) is the diffraction
intensity of light double-slit diffraction including decoherence
effects, and Eq. (43) is the diffraction intensity of light
double-slit diffraction considering coherence superposition.
\vskip 8pt {\bf

5. Numerical result} \vskip 8pt

In this section, we report our numerical results of diffraction
intensity for light double-slit diffraction. The theory result of
quantum theory is from Eq. (45), and Eq. (47) is the theory result
of classical electromagnetic theory. The Ref. [20] is the light
double-slit diffraction experiment. In [20], two slit width are
$a=1.3\times 10^{-4}m$, the distance between the two slit
$d=4\times 10^{-4}m$, slit to the screen distance $l=4 m$, and the
wavelength of the light $\lambda=916\times 10^{-9}m$. From FIG. 2,
because $l\gg a+d$, we have $\beta_1\approx\beta_2=\beta$. In our
calculation, we take the same experiment parameters above. The
theory parameters are taken as: the slit length $b=4.4\times
10^{-3}m$, slit thickness $c=8.5\times 10^{-5}m$, $\alpha=0$,
$A_{1j}=160.9$, $A_{2j}=159.3$, $c_{1}=0.715$, $c_{2}=0.699$
$(|c_{1}^{2}|+|c_{2}^{2}|=1)$ and the quantum coherence degree
$\nu=0.873$. For the classical electromagnetic theory, the
double-slit diffraction intensity is
\begin{eqnarray}
I=4I_{0}\frac{\sin^{^{2}}(\frac{\pi a
\sin\beta}{\lambda})}{(\frac{\pi a
\sin\beta}{\lambda})^{2}}\cdot\cos^{^{2}}(\frac{\pi d
\sin\beta}{\lambda}).
\end{eqnarray}
In FIG. 3, the point is the experimental data from Ref. [20]. The
solid curve is the calculation result of quantum theory from Eq.
(45), which include decoherence effects. We can find the quantum
calculate results is in accordance with the experiment data. In
Fig. 4, the point is the experimental data from Ref. [20]. The
solid curve is the calculation result of classical theory from Eq.
(47). We also find the theory results of classical electromagnetic
have a certain deviation with the experimental data. The deviation
mainly come from: (1) The theory curve intersect at the abscissa
axis $\beta$, but experiment values have not intersection point
with axis $\beta$. (2) The maximum values of calculation are less
than the experimental date. So, the classical electromagnetic
theory is an approximate approach to study light diffraction, and
the more accurately approach is the quantum theory of light.
\begin{figure}[tbp]
\includegraphics[width=8.5 cm]{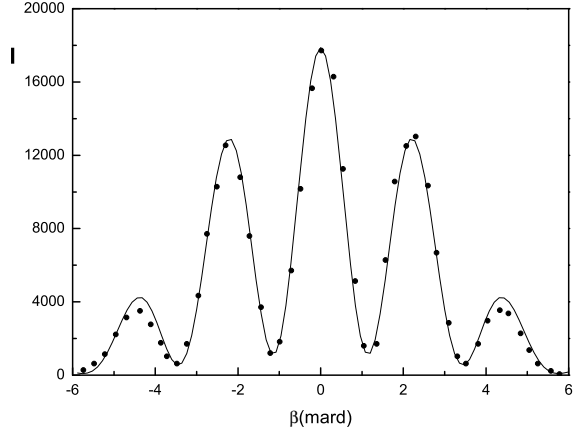}
\caption{Comparing the calculation result of quantum theory with
the experiment data}
\end{figure}
\begin{figure}[tbp]
\includegraphics[width=8.5 cm]{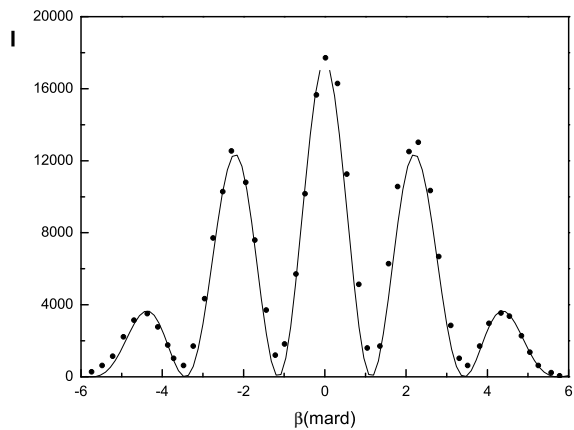}
\caption{Comparing the calculation result of classical theory with
the experiment data }
\end{figure}
\newpage
 \vskip 10pt
{\bf 6. Conclusion} \vskip 8pt

In conclusion, we have studied double-slit diffraction of light
with the approaches of quantum theory and classical
electromagnetic theory. In quantum theory, we give the relation
among diffraction intensity and slit length, slit width, slit
thickness, wave length of light and diffraction angle. In
classical electromagnetic theory, only give the relation among
diffraction intensity and slit width, wave length of light and
diffraction angle. Obviously, the quantum theory include more
diffraction information than the classical electromagnetic theory.
By calculation, we find the classical electromagnetic theory
result has a certain deviation with the experimental data, but the
quantum calculate result is in accordance with the experiment
data. So, the classical electromagnetic theory is an approximate
approach, and the quantum theory is more accurately approach for
studying light diffraction.
 \\

\end{document}